\documentclass[pra,showpacs,floatfix,amsmath,amsfonts,twocolumn]{revtex4}
\usepackage{dcolumn}
\usepackage{bm}
\usepackage{epsfig}
\usepackage{color}
\usepackage{graphics}
\usepackage{amssymb}

\begin{document}

\title{Flat bands and dynamical localization of spin-orbit coupled Bose-Einstein condensates}

\author{Fatkhulla Kh. Abdullaev}
\affiliation{Physical-Technical Institute, Uzbekistan Academy of Sciences, Tashkent}
\author{Mario Salerno}
\affiliation{Dipartimento di Fisica ``E.R. Caianiello'', and Istituto Nazionale di Fisica Nucleare - Gruppo Collegato di Salerno, Universit\'a di Salerno, Via Giovanni
Paolo II, 84084 Fisciano (SA), Italy}
\begin{abstract}
Flat bands and dynamical localization of binary mixtures of Bose-Einstein condensates, with spin-orbit
coupling subjected to a deep optical lattice which is shaking in time and to a periodic time modulation of the Zeeman field, are investigated. In contrast with usual dynamical localization in the absence of spin-orbit coupling, we find that to fully suppress the tunneling in the system the optical lattice shaking is not enough, and
a proper tuning of the spin-orbit term, achievable via the Zeeman field modulation, is also required. This leads to a sequence of Zeeman parameter values where energy bands become flat, the tunneling in the system is suppressed, and the dynamical localization phenomenon occurs. Exact wave functions at the dynamical localization points
show that the binary mixture localizes on a dimer with the two components occupying different sites. This type of localization occurs in exact form also for the ground state of the system at the dynamical localization points
in the presence of nonlinearity and remains valid, although in approximate form, for a wide range of the Zeeman parameter around these points. The possibility of observing the above phenomena in real experiments is also briefly discussed.
\end{abstract}
\maketitle

\section{Introduction}

Significant attention is presently devoted to the study of
time-periodically driven many-body systems \cite{rev1,rev2} that exhibit interesting transport phenomena that resemble the ones observed in condensed matter physics under the action of static or  time-periodic electric fields. In this context, it is well known
that a constant electric field cannot induce transport in perfect
crystals, due to the phenomenon of Bloch oscillations. In the
presence of a time-periodic electric field, however, transport
becomes generically possible, except for specific ratios of the
field amplitude and frequency for which the phenomenon of
dynamical localization (DL) appears.

As  first demonstrated in Ref.\cite{DK} for the Schr\"odinger  tight-binding  model  of electrons in perfect crystals, DL emerges due to  the tunneling suppression between adjacent sites induced  by the periodic electric field.   For harmonic fields this happens when the amplitude frequency ratio matches zeros of the Bessel function $J_0$. DL is not a peculiarity of  linear systems  but exists also in the presence of nonlinearity (interactions)  as has been shown for the discrete nonlinear Schr\"odinger equation (DNLS)  in Ref.\cite{cai}  and  its quantum version (Bose-Hubbard model) in Ref.\cite{holthaus}. To date, DL has been observed in many physical systems among which spin chains \cite{rev1}, periodically curved arrays of optical waveguides \cite{Longhi,Sterke}, cold atoms loaded in shaken optical lattices \cite{Lignier}.

In Bose-Einstein (BEC) condensates, the analogues of periodic electric fields can be realized  by means of shaking optical lattices. This  leads to very interesting phenomena including the generation of synthetic gauge fields \cite{Struck}, topological insulators
\cite{Hauke}, etc. In these systems it is also possible to modulate the interactions (scattering lengths) in time, a fact that allows  to change the inter-site tunneling of BEC in optical lattices in a manner that depends not only on the amplitude and frequency of the modulation but also on the relative atomic imbalance between adjacent sites \cite{Haggi,AKS}. This leads to the appearance of new quantum phases \cite{Santos}, density dependent gauge fields \cite{Greischner}, matter wave excitations localized on a compact domain (compactons) \cite{AKS, ASKA}. Time-dependent modulations of the scattering lengths  have been also shown to be very effective to suppress dynamical instabilities and to induce long-living Bloch oscillations \cite{SKB} and DL \cite{BKS} of matter wave  gap-solitons.

Spin-orbit coupling (SOC)  opens new possibilities for investigating the above phenomena in BEC systems. In particular,   due to the  interplay between SOC, periodicity   and nonlinearity, DL could display interesting new features. In BEC systems the effective SOC stemming from internal atomic states  which are coupled  by Raman  laser fields \cite{Galitskii-Nature}   can be tuned by means of   fast and coherent modulations of the laser intensities  \cite{Zhang}. This can be achieved via modulations of the Raman term, as experimentally  demonstrated in Ref.\cite{Garcia}, by modulating gradient magnetic fields \cite{Luo}, or  by  time periodic modulations of the Zeeman field \cite{SAGT}.

In spite of this, however, very few investigations of flat bands and DL for SOC-BEC systems presently exist. In this context we mention the spin-dependent DL of a SOC single atom in a driven optical bipartite lattice \cite{Luo2016}, the  DL in a SOC  two-level atom  trapped in periodic potential under the action of weak harmonically varying linear force \cite{KKZT},  and the  dynamical suppression of the tunneling in a double well potential for a non interacting (linear) SOC-BEC system \cite{KKV}. Moreover, almost no  studies  exist on the effect of combined modulations on the DL phenomenon. In this respect  we can mention only the work \cite{Greischner2}, where combined modulations of interactions and lattice shaking are used to generate the extended  Hubbard models with asymmetric hopping which predict new quantum phases in BEC.

The  aim of the present paper  is to investigate DL phenomena in binary BEC mixtures subjected to optical lattice shaking,  time periodic Zeeman field, and equal SOC contributions of Rashba and Dresselhaus type. For this we use a  tight binding model for  BEC-SOC mixture appropriate for  deep optical lattices \cite{SA,BGPMHM}  and treat the time modulations in the fast frequency limit. This leads  to an effective time-averaged Hamiltonian system which can be  analytically solved in the linear case  and analytically (at DL points) and numerically investigated in the nonlinear case.

As a result we find  that, in contrast with  usual DL (e.g. in absence of SOC), the shaking of the optical lattice  alone is not enough to fully suppress the tunneling, but suitable tunings of the SOC term  with the optical lattice shaking, achieved via  Zeeman field modulation, are also required. This leads to a sequence of Zeeman parameter values  for which DL can occur (DL points). We show that at DL points the energy bands become flat and the tunneling  is fully suppressed. In the linear case, exact wave functions derived at the DL points show that the localization occurs on a dimer with the BEC components occupying different sites.
We show that this holds true, in exact form, also for the ground state wavefunctions at the DL points   in the presence of intra-species contact interactions and remains valid, in approximate form, for a wide range around these points.

The paper is organized as follows. In Section II we introduce the  model equations and derive  the effective averaged Hamiltonian system. In section III we study the dispersion relation and the linear spectrum as a function of the system parameters. In Section IV exact analytical   wavefunctions at the DL points of the linear case are derived and in Section V we extend results to the ground state of the system in the presence of nonlinearity. Finally, in section VI  we discuss parameters design for possible experiments and briefly summarize the main results of the paper.

\section{Model and averaged equations}

A BEC with equal Rashba and Dresselhaus SOC contributions loaded in a deep optical lattice  can be described in the mean field approximation by the following
DNLS equation~\cite{SA, BGPMHM}:
\begin{equation}
\begin{split}
i\frac{d u_n}{dt}=&\,-\Gamma(u_{n+1}+u_{n-1})+ i\frac{\sigma}{2}(v_{n+1}-v_{n-1})\\
&\,+\Omega u_n+ (\gamma_1 |u_n|^2 + \gamma  |v_n|^2) u_n + f(t) n u_n, \\
i \frac{d v_n}{dt}=&\, -\Gamma (v_{n+1}+v_{n-1}) + i\frac{\sigma}{2} (u_{n+1}-u_{n-1})\\
&\,- \Omega v_n + (\gamma |u_n|^2 + \gamma_2 |v_n|^2) v_n + f(t) n v_n .
\end{split}
\label{eq1}
\end{equation}
Here $\sigma$ and $\Omega$ denote the SOC and the Zeeman parameters while the linear ramp potential, modeling  the optical lattice shaking,  is assumed to be time-periodic with amplitude  $f(t)=f_0 \cos(\omega t)$. In order to have SOC tunability we also assume that the Zeeman term is varying periodically in time as $\Omega=\Omega(t)=\Omega_0 + \Omega_1 \cos(\omega t)$, where $\Omega_0$ is a constant fixed part and $\Omega_1$ is the amplitude of the part that is modulated with the same frequency  $\omega$ of the lattice shaking. In the following we assume rapid and strongly varying fields, $\Omega(t),  f(t)$, of the form
\begin{equation}\label{eq2}
\Omega(t)=\Omega_0 + \frac{1}{\epsilon}\Omega_1 \cos(\omega\, \tau),\;\; f(t)=\frac{1}{\epsilon} f_0 \cos (\omega\, \tau),
\end{equation}
with $\epsilon \ll 1$ and $\tau=\frac{t}{\epsilon}$ denoting a fast time variable.
To remove the explicit fast time dependence from Eq. (\ref{eq1}) it is convenient to perform the following transformation
\begin{equation}\label{eq3}
u_n=U_n e^{-i\frac{\sin(\omega t)}{\omega}(\Omega_1 + \gamma_0 n)},
v_n=V_n e^{i\frac{\sin(\omega t)}{\omega}(\Omega_1 - \gamma_0 n)}.
\end{equation}
By substituting into Eq.(\ref{eq1}) and averaging with respect to the fast time variable we obtain the following averaged system:
\begin{eqnarray} \label{avGP}
i \dot U_{n} = &-&\Gamma J_0(\chi)(U_{n+1}+U_{n-1}) + \frac{i\sigma}{2}(J_0^{-} V_{n+1}-J_0^{+} V_{n-1}) \nonumber\\ &+& \Omega_0 U_n + (\gamma_1 |U_n|^2 + \gamma |V_n|^2) U_n,\\
i \dot V_{n} = &-&\Gamma J_0(\chi)(V_{n+1}+V_{n-1}) + \frac{i\sigma}{2}(J_0^{+} U_{n+1}-J_0^{-} U_{n-1}) \nonumber \\ &-& \Omega_0 V_n + (\gamma_2 |V_n|^2 + \gamma |U_n|^2) V_n,\nonumber
\end{eqnarray}
where $J_0(\chi)$ denotes the zero-order Bessel function of a variable $\chi$, while $J_0^\pm$ stands for
\begin{equation}
J_0^\pm\equiv J_0^\pm (\eta)=J_0(\eta \pm \chi)
\label{J0pm}
\end{equation}
with
$$
\chi=\frac{f_0}{\omega}\;\;\;\; , \;\;\;\; \eta =2 \,\frac{\Omega_1}{\omega}.
$$

Notice that Eq. (\ref{avGP}) has two conserved quantities: the norm
\begin{equation}
N= \sum_n (|U_n|^2 + |V_n|^2)
\end{equation}
and the Hamiltonian (energy)
\begin{eqnarray}
&& H = \sum_n \left\{ - J_0(\chi) \Gamma(U_n^* U_{n+1} + V_n^* V_{n+1})\; + \right. \nonumber
\\&& \left. \;\;\;\;\; i \frac{\sigma}{2} U_n^{*}(J_0^- V_{n+1} - J_0^+ V_{n-1})\; + \right. \nonumber
\\&& \left. \;\;\;\;\; \frac{\Omega_0}{2}(|U_n|^2-|V_n|^2) + \; c.c. \right\} + E_{int}, \nonumber
\label{Energy}
\end{eqnarray}
where $E_{int}$ is the interaction energy,
\begin{equation}
E_{int}=\sum_n\{\frac{1}{2}(\gamma_1|U_n|^4+\gamma_2|V_n|^4)+\gamma |U_n|^2 |V_n|^2\} ,
\label{Eint}
\end{equation}
and c.c. denotes the complex-conjugate of the expression in the curly bracket.
From Eq.(\ref{avGP}) it is clear that the inter-well tunneling induced by the dispersive term $\Gamma$ can be suppressed if $\chi$ is taken as a zero of the Bessel function $J_0$. Notice that  in absence of SOC this would be sufficient to fully suppress the tunneling in the system  but in the presence of SOC this is not so because tunneling remains possible through the SOC (see the $\sigma$  term in Eq. (\ref{avGP})).
\begin{figure}[t]
\centerline{
\includegraphics[width=6.5cm]{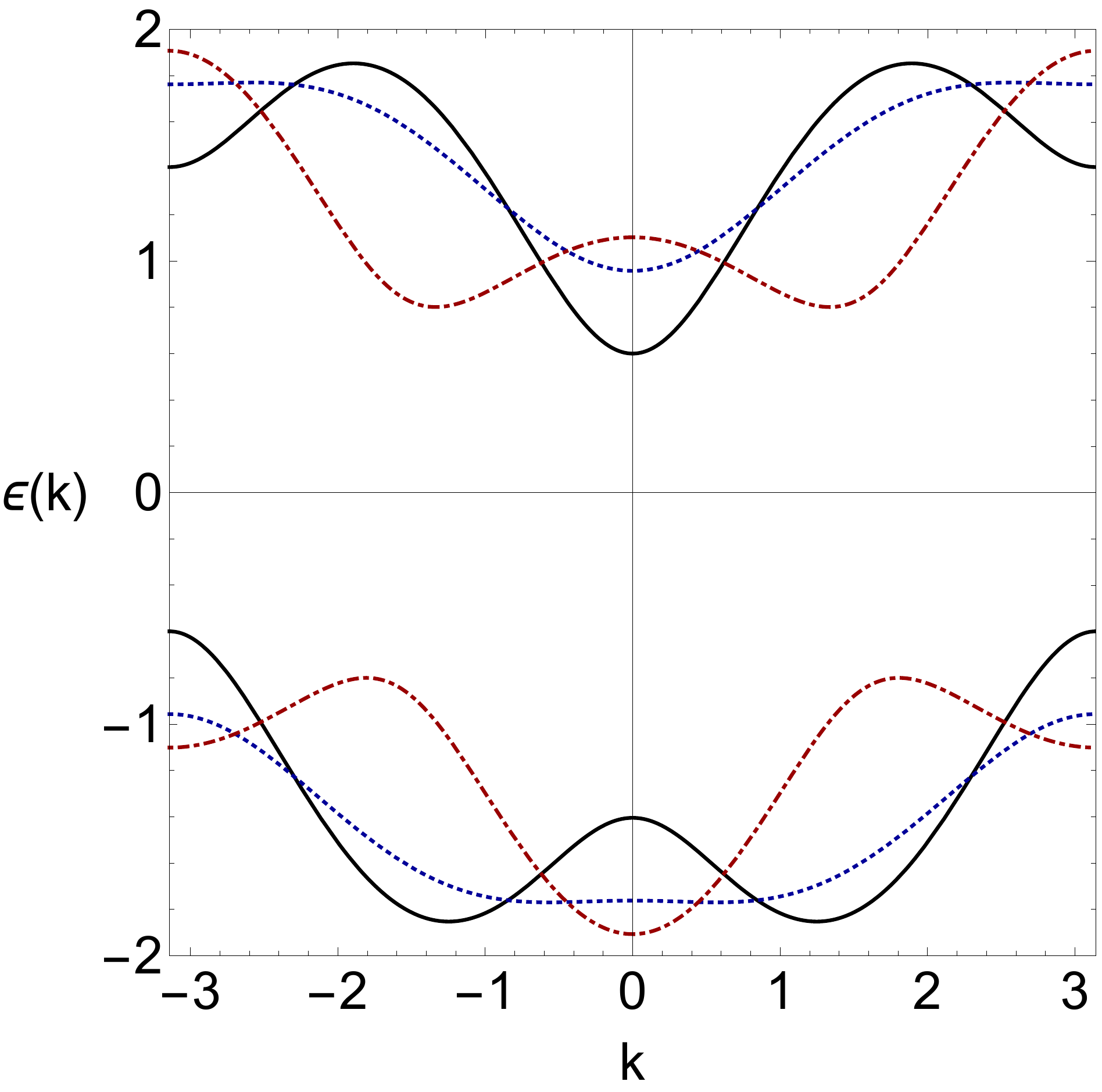}
}
\centerline{
\includegraphics[width=6.5cm]{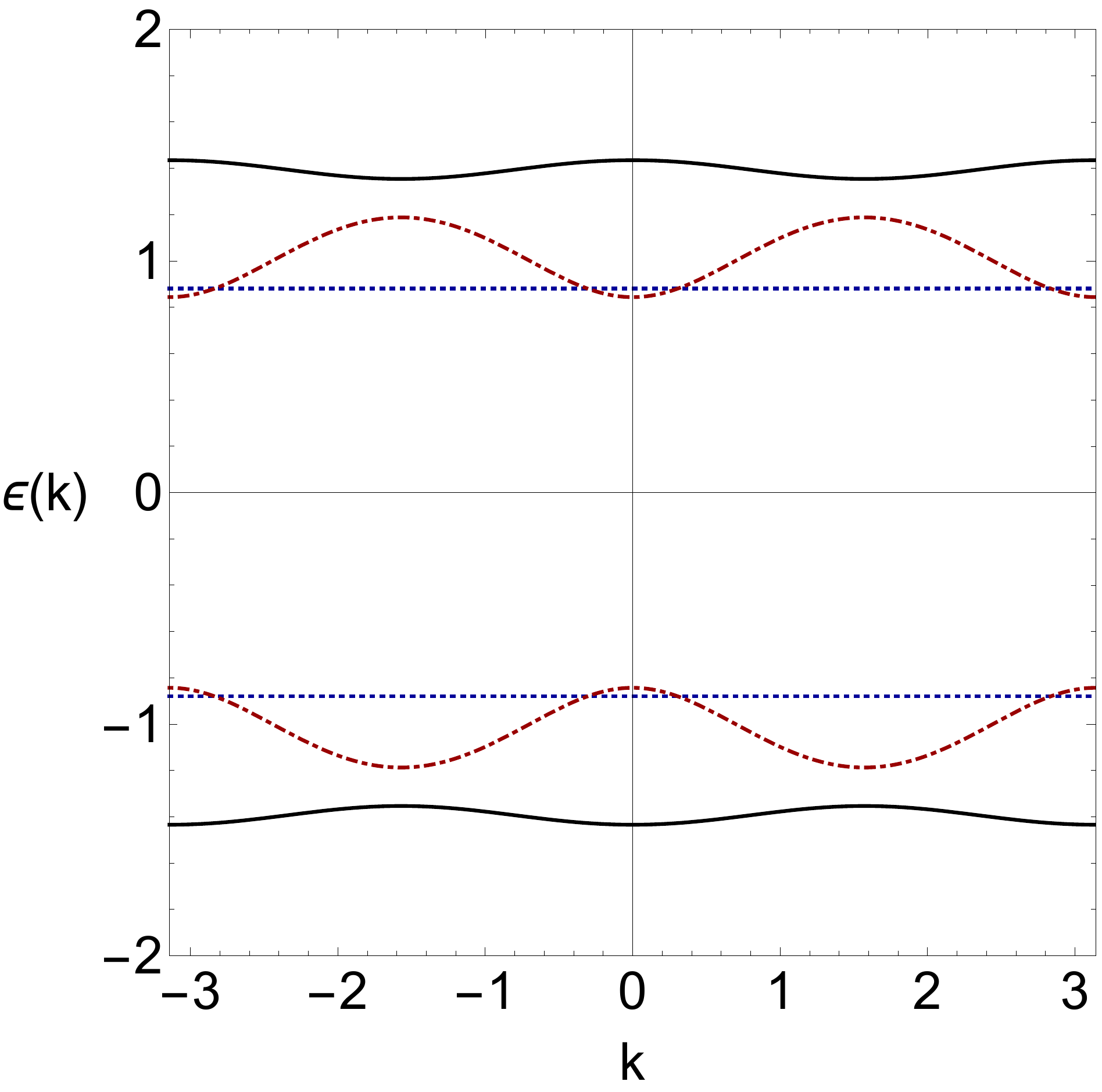}
}
\vspace{-0.cm}
\caption{Typical dispersion curves of the linear SOC-DNLS system ($g=0$) for two values of the optical lattice shaking parameter: $\chi=1.2$ (top panel), $\chi=\xi_1=2.40483$ (bottom panel) and different Zeeman field modulations. The black continuous, blue dotted and red dot-dashed curves in the top and bottom  panels refer to $\eta=0.5, 1.0, 2.5$, and $\eta =3.0, 4.80965, 5.0$, respectively. Other parameters are fixed as $\Gamma=0.3, \Omega_0= 0.8, \sigma=2.5$.
}
\label{fig1}
\end{figure}

\section{Dispersion relations and linear energy spectrum}
In the absence of contact interactions [$\gamma=\gamma_1=\gamma_2=0$ in Eq. (\ref{avGP})] it is possible to derive the dispersion relation by assuming a dependence of $U_n, V_n$ on time and on the lattice site $n$ of the form
\begin{equation}
U_n(t)= A \exp{i (k n - \epsilon t)},\;\; V_n(t)= B \exp{i (k n - \epsilon t)},
\end{equation}
where $A, B$ are real constants, $k$ is the crystal momentum varying in the first Brillouin zone,
$k \in [-\pi,\pi]$, and  $\epsilon$ has the physical  meaning  of chemical potential ($\equiv$ energy  in linear case). The dispersion relation, e.g.  the dependence of $\epsilon$ on $k$, directly  follows from the compatibility condition of the resulting linear system, and one can easily show that it leads to
\begin{eqnarray}
&& \epsilon_\nu (k)= - 2 \Gamma J_0(\chi) \cos(k) + \\
&& \nu \sqrt{\Omega_0^2+\frac{\sigma^2}{4} (J_0^{+}(\eta)-J_0^{-}(\eta))^2+
\sigma^2 J_0^{+}(\eta) J_0^{-}(\eta) \sin(k)^2} \nonumber
\label{disprel}
\end{eqnarray}
with the index $\nu$ assuming the values $\nu=-1, 1$, in correspondence of the lower and upper branches of the dispersion curve, respectively.

Notice that in the absence of modulations (e.g. $f_0=\Omega_1=\eta=\chi=0$) we have $J_0(\eta)=J_0^\pm=1$ and the above dispersion relation reduces the the one in Ref. \cite{SA} for the case of a static optical lattice and constant Zeeman field. Similarly, in absence of the shaking of the optical lattice (e.g. for $f_0=\chi=0$), Eq. (\ref{disprel}) reproduces the one considered in Ref. \cite{SAGT} for the case of SOC tunability induced by time dependent Zeeman fields.
Typical dispersion curves for different modulating parameter values are depicted in Fig.\ref{fig1}. Notice from the bottom  panel the occurrence of a flat band for the value  $\eta= 4.80965$ which is related to a zero of the Bessel function $J_0$ as we shall see in the following.

\section{SOC dynamical localization: linear case}

In this section we consider the effects of  the optical lattice shaking and Zeeman modulation on the band flatness,  suppression of tunneling and DL existence, in the absence of  any contact interaction. To this end  we fix the optical lattice shaking parameter $\chi$ to a zero of the Bessel function, say $\chi=\bar \xi$, so that the effective inter-well tunneling constant, $J_0(\chi) \Gamma$, vanishes. In this case the lower and upper bands are  related by the symmetry $\epsilon_{1}(k)=-\epsilon_{-1}(k)$, and their  dependence on $k$ is fully controlled by  the Zeeman parameter $\eta$ through the factor $ \sigma^2 J_0^+  J_0^-$ in Eq.(\ref{disprel}).
\begin{figure}[t]
\vspace{-1.cm}
\centerline{
\includegraphics[width=8.5cm]{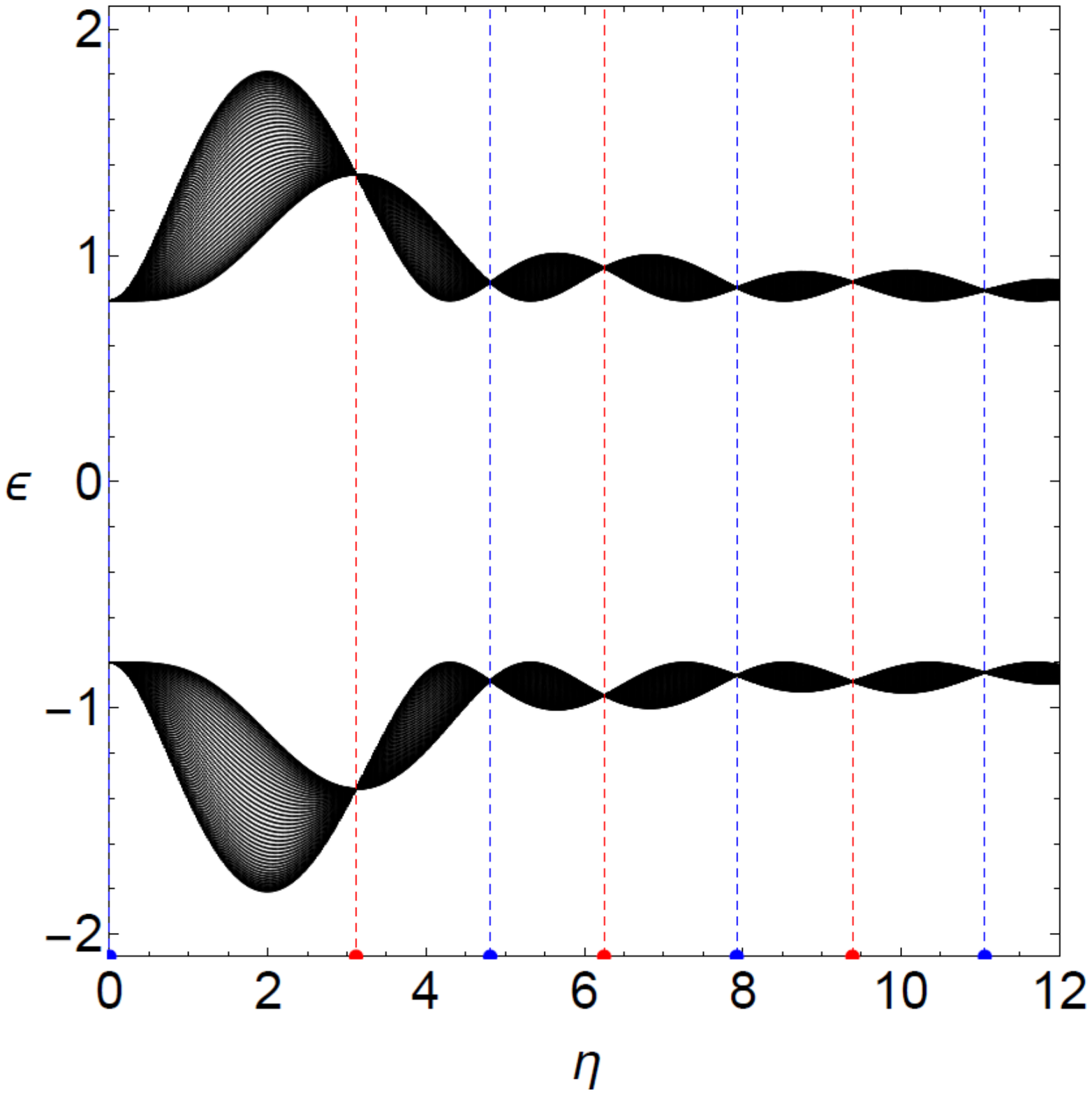}
}
\vskip -4.2cm
\centerline{
\includegraphics[width=8.5cm]{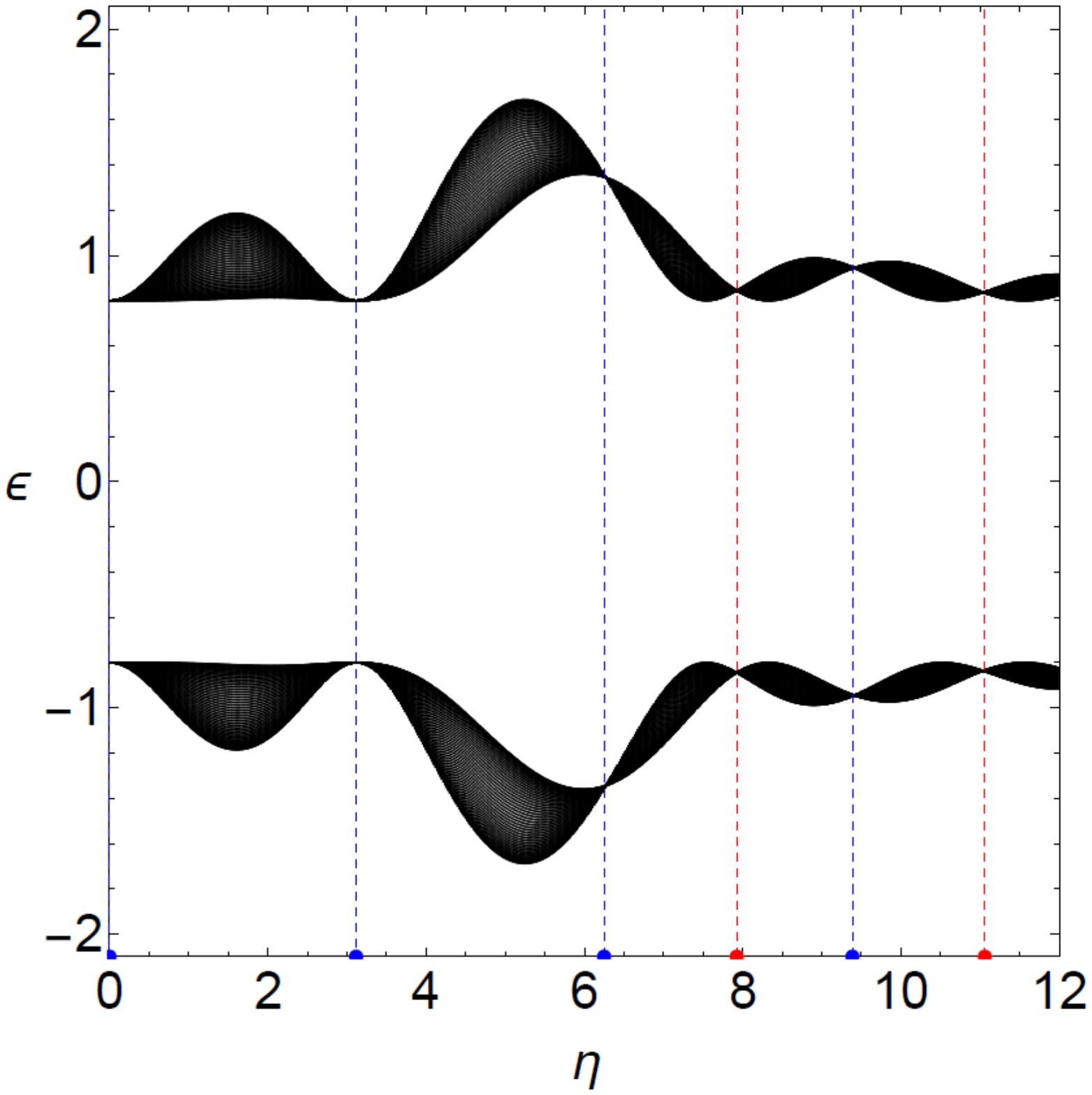}
}
\vskip -4.2cm
\caption{Energy spectrum versus $\eta$ for parameter $\chi$ fixed as the first (top panel) and second (bottom panel) zero of the Bessel function $J_0$, e.g. $\chi=2.40483$ and $\chi=5.52008$, respectively. Blue and red dots correspond to the $\eta^-$ and $\eta^+$ values (e.g. to solutions  of the equations $J^+=0$ and $J^-=0$, respectively), leading to the sequences in Eq. (\ref{fbseq}): $\{0, 3.11525, 4.80965, 6.2489, 7.9249, 9.38671, 11.0586\}$ (top panel) and $\{0., 3.13365, 6.27146, 7.9249, 9.41084, 11.0402\}$ (bottom panel). Dashed blue and red lines have been drawn just as a guide for the eyes through the band shrinking points. The full spectrum for a given $\eta$ has been derived for a chain of $99$ sites. Other parameters are fixed as in Fig.\ref{fig1}.
}
\label{fig2}
\end{figure}

Considering  the SOC parameter  different from zero, we have that the band  flatness  is achieved  when one of the equations $J_0^\pm(\eta) \equiv J_0(\eta \pm \bar\xi)=0$  is satisfied. This occurs for $\eta$ taken as:
 \begin{equation}
 \eta^\pm_n=\xi_n\pm\bar\xi,
 \label{etapm}
\end{equation}
with $\xi_n$, $n=1,2,...$, denoting the n-th zero of $J_0$. Flat bands in $k$-space are found in correspondence of the N-fold degenerate eigenvalues
\begin{equation}
\epsilon_\nu= \nu \sqrt{\Omega_0^2 + \frac{\sigma^2}{4} J_0^\pm(\eta_n^\pm)^2},
\label{flatbandenergy}
\end{equation}
which depend on $\sigma$ and with $\nu=-1, 1,$ referring to the lower and upper bands, respectively. From this it follows that for $\chi=\bar\xi \equiv \xi_m$ the $m$-th zero of $J_0$, the band flatness occur in correspondence of the sequence of $\eta$ values (DL points), listed in increasing order,
\begin{equation}
 \{\eta^-_m, \eta^-_{m+1}, ... , \eta^-_{2m}, \eta^+_1, \eta^-_{2m+1}, \eta_2^+, \eta_{2m+2}^-,  \eta_3^+,\eta_{2m+3}^-, ... \}.
\label{fbseq}
\end{equation}
Thus, for $m=1$, e.g. $\chi=\xi_1=2.40483$, the first zero of $J_0$, the $\eta$-sequence is
\begin{equation}
\{\eta^-_1,\eta^-_2,\eta^+_1,\eta^-_3,\eta^+_2,\eta^-_4,\eta^+_3,\eta^-_5,\eta^+_4,\eta^-_6, ,\eta^+_5,\eta^-_7,\;...\;\}.
\label{seq1}
\end{equation}
Notice that the flat band depicted in the bottom panel of  Fig.\ref{fig1} just corresponds to the $\eta_1^+$ value in the above  sequence. Also notice that at DL points the band velocity $v=d\epsilon(k)/dk$ vanishes for all values $k$, so the transport is fully suppressed and DL occurs.
\begin{figure}[t]
\vspace{-1.cm}
\centerline{
\includegraphics[width=8.5cm]{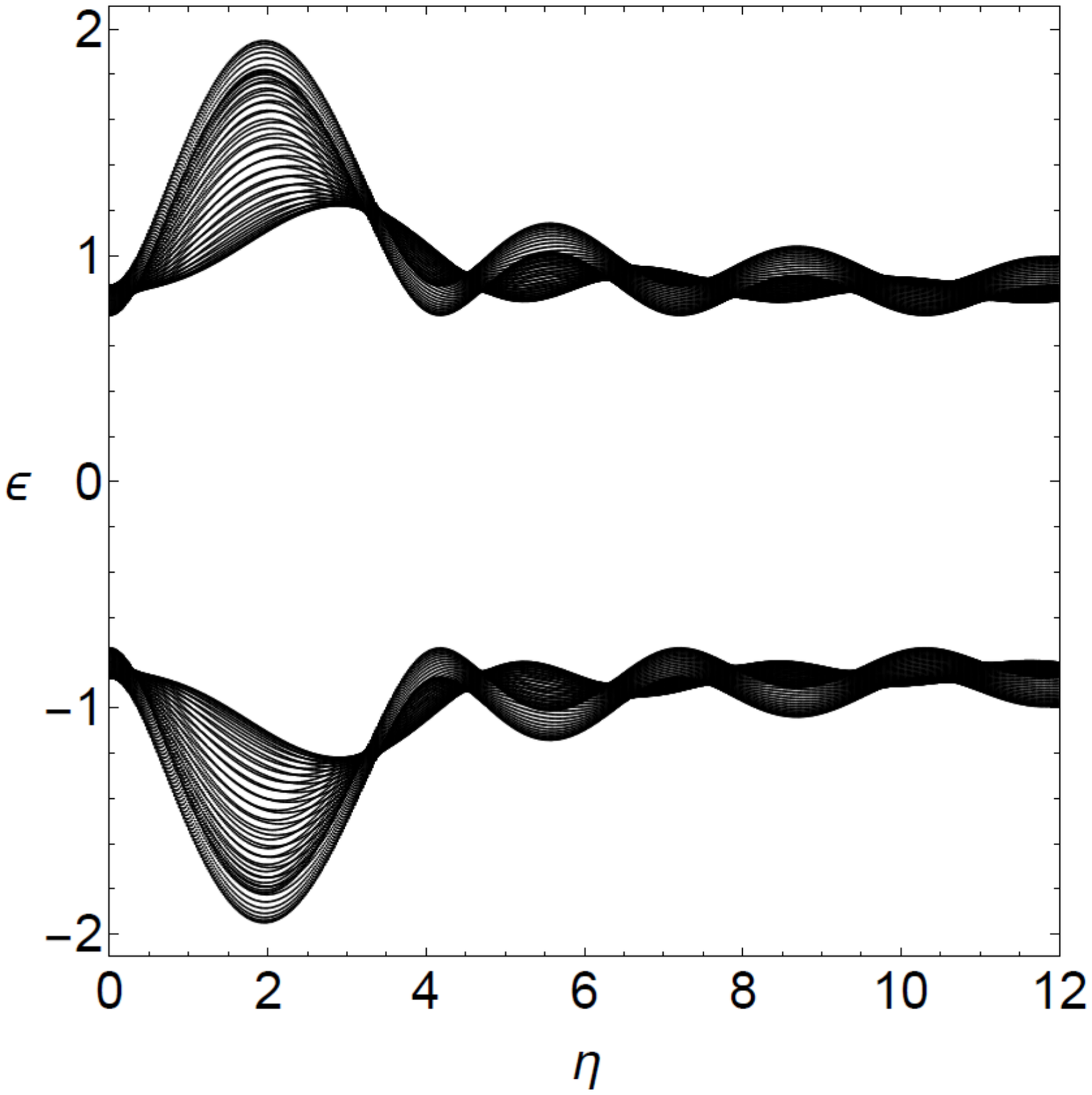}
}
\vskip -4.2cm
\centerline{
\includegraphics[width=8.5cm]{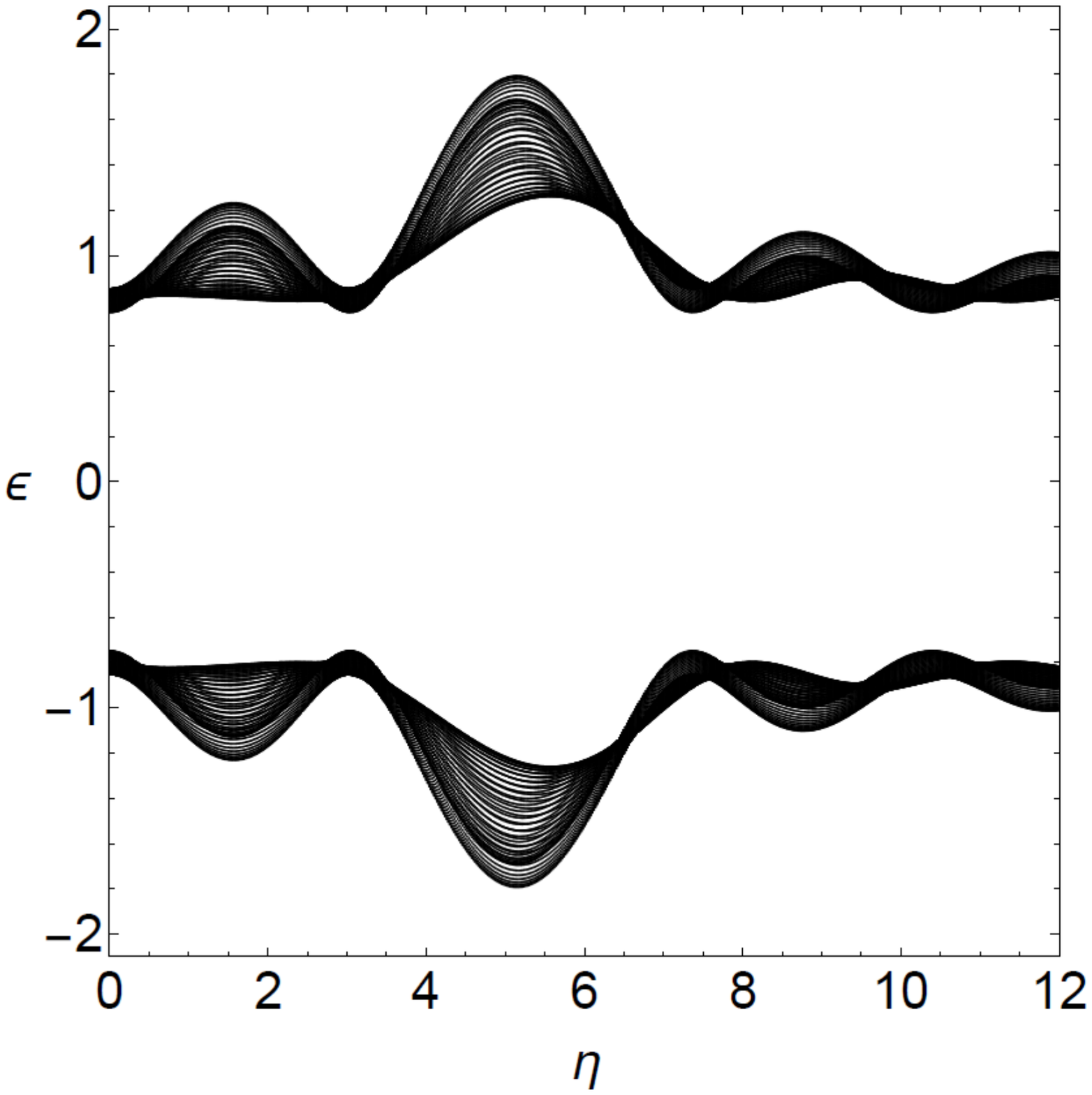}
}
\vskip -4.2cm
\caption{Energy spectrum versus $\eta$ as in Fig.\ref{fig2} but for $\chi$ detuned from the first two zeros of $J_0$  by $-0.25$, e.g. $\chi=\xi_i - 0.25$ with $i=1,2$ for top and bottom panel respectively. Other parameters are fixed as in Fig.\ref{fig2}.
}
\label{fig2bis}
\end{figure}

In Fig. \ref{fig2} we report  the energy spectrum as a function of the  modulation parameter $\eta$ for $\chi$ fixed as the first (top panel) and second (bottom panel) zero of $J_0$. In both cases the effective interwell tunneling $J_0(\chi)\Gamma$ vanishes and we see the shrinking of the spectrum into two degenerated points  exactly at the values predicted by Eq. (\ref{fbseq}). Notice that the degenerate energies at these points correspond to different values of the crystal momentum $k$ and generate the upper and lower flat seen in the  bottom panel of Fig.\ref{fig1}, when plotted in the $k$-space.

In Fig.\ref{fig2bis} we show the energy spectrum as in Fig.\ref{fig2} but for $\chi$ slightly different from a zero of $J_0$. Since the effective interwell tunneling is not zero, no shrinking of the spectrum into single points can be observed in this case. Obviously, no flat bands, tunneling suppression or DL phenomena can arise for any value of $\eta$.

Notice that when the Zeeman field modulation is switched off, e.g. $\eta=0$, the flatness of the bands and the DL phenomenon occur when the intrawell tunneling is suppressed, e.g. when $\chi$ matches a zero of $J_0$. In this case, however, $J_0^\pm = J_0(\chi)=0$ so there are only  the trivial flat bands $\epsilon_\nu=\nu \Omega_0$ and no SOC contribution to the DL at all, since  $\sigma$  disappears from the dispersion relation.
From this it is clear  that in the presence of SOC the suppression of the interwell tunneling by means of the optical lattice shaking alone is not enough to induce DL, and  the matching of the Zeeman field parameters with the values in Eq. (\ref{fbseq})  is absolutely necessary for DL and band flatness to exist.

Let us now investigate the eigenstates of the  system at the DL points. In this respect, we remark from  Eq. (\ref{avGP}) that for  $J_0(\chi)=0$ and $\eta$ satisfying  one of the two equations:  $J_0^\pm(\eta)=0$, the system reduces to  a chain  of uncoupled dimers  described by the equations
\begin{equation}
i \dot A \mp \frac{i \sigma}{2}J_0^{\mp} B -\Omega_0 A=0,\;\;\;\;\; i \dot B \pm \frac{i \sigma}{2}J_0^{\mp} A +\Omega_0 B=0,
\end{equation}
with  $A, B$ standing for $U_n, V_{n \pm 1}$ and the signs appearing in the equations related to which of the two equations $J_0^\pm=0$ is satisfied by $\eta$. In this case one can readily check that the following stationary dimer solution exists:
\begin{eqnarray}
&& A_{\nu}\equiv U_{n}=-2 i \sqrt{\frac{(\epsilon_{\nu}+\Omega_0)(\epsilon_{\nu}^2-\Omega_0^2)}{2 \epsilon_{\nu} (\sigma J_0^{\mp})^2}}\; e^{-i \epsilon_{\nu} t} \delta_{n,n_0}\;,  \\
&&B_{\nu}\equiv V_{n}= - \nu \sqrt{\frac{\epsilon_{\nu} -\Omega_0}{2 \epsilon_{\nu}}}\; e^{-i \epsilon_{\nu} t} \delta_{n, n_0 \pm 1}\; ,
\end{eqnarray}
with $\epsilon_{\nu}$ being the flat band energy at the DL point in Eq.(\ref{flatbandenergy}).
\begin{figure}[t]
\vspace{-1.cm}
\centerline{
\includegraphics[width=8.cm]{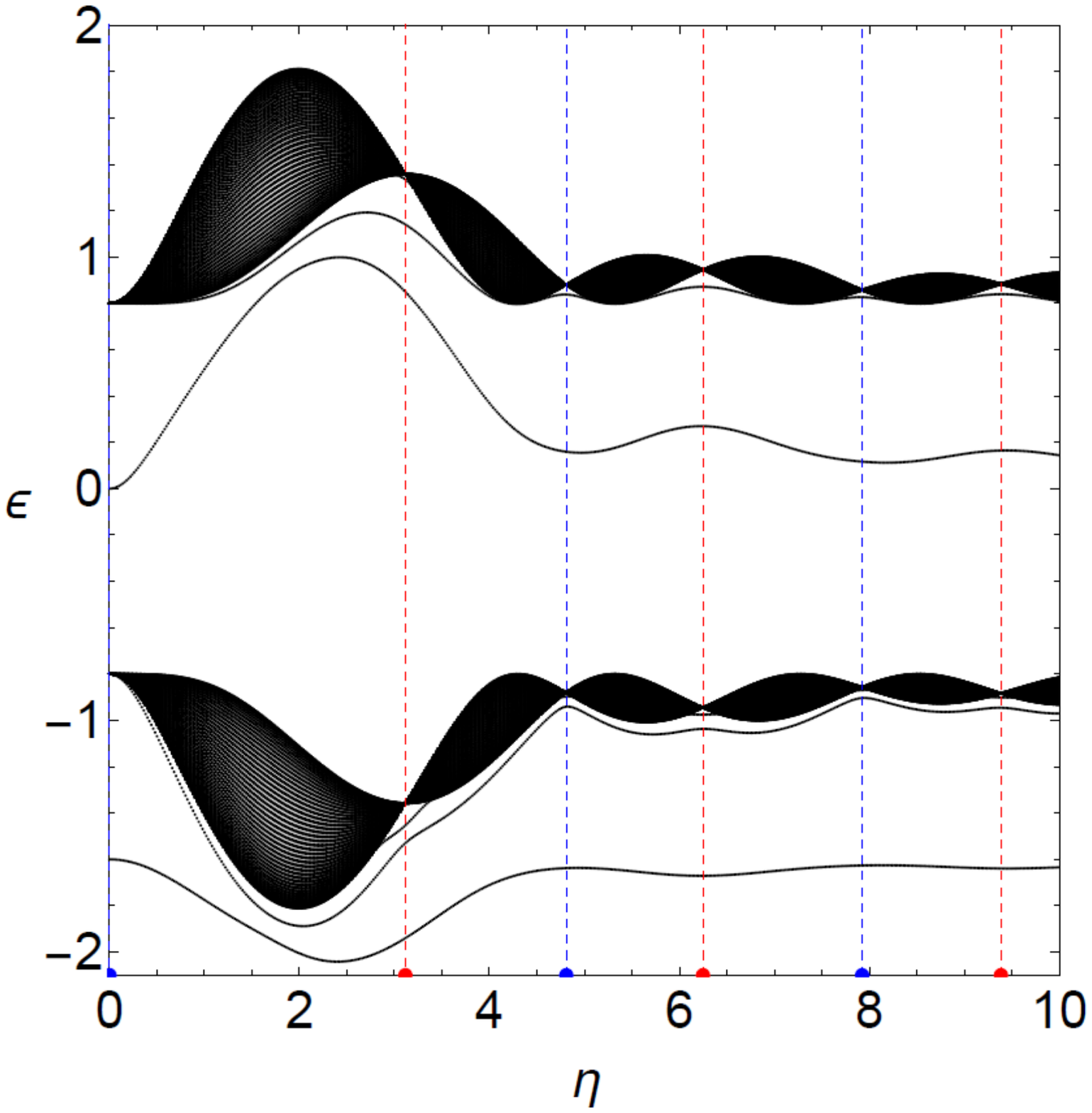}
}
\vskip -5.2cm
\centerline{
\includegraphics[width=8.cm]{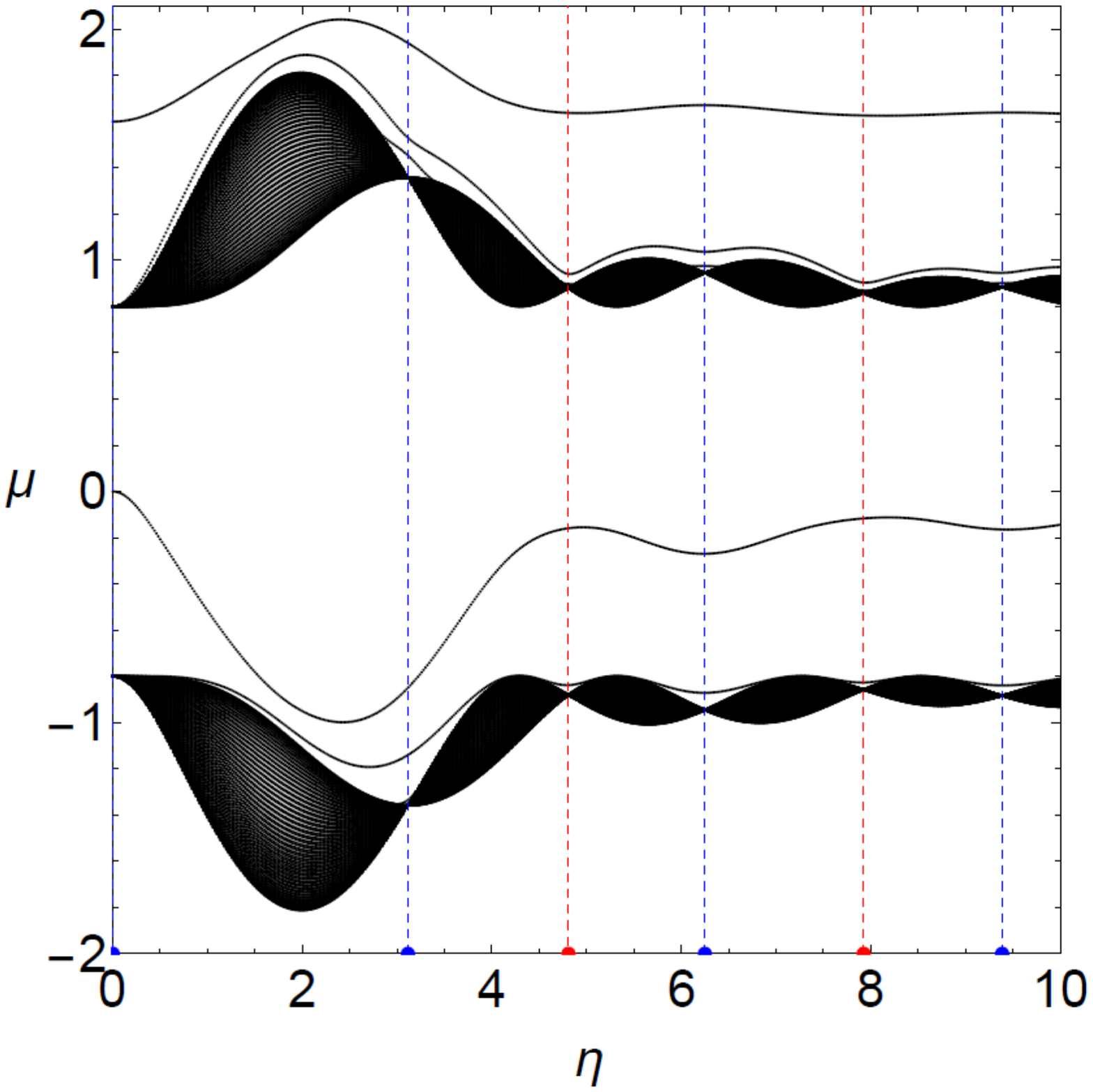}
}
\vspace{-3.5cm}
\caption{Chemical potential $\mu=\epsilon+E_{int}$ for the equally attractive case: $\gamma_1=\gamma_2=\gamma=-0.8$ (top panel) and the equally repulsive case: $\gamma_1=\gamma_2=\gamma=0.8$ (bottom panel). Other parameters are fixed as $\chi=\xi_1=2.40483, \Gamma=0.3, \Omega_0=0.8, \sigma=2.5$.
}
\label{fig3}
\end{figure}

Notice that the above  eigenstates are  normalized  according to $|A|^2 + |B|^2=1$ and they  can exist, on any lattice point, for a total of $N$ orthonormal states per band ($N$ being the number of lattice sites in the chain). Actually, from this complete  set of localized states one can construct extended (Bloch) states at the DL points in terms of a Fourier transform. The fact that the two components of the above dimer solution are localized on different sites implies that they exist also  in the presence of the intra-species interactions, e.g. at the DL points the SOC system remains dimerized even in the presence of the inter-species nonlinearity if the nonlinearity  intrasite scattering lengths are detuned to zero.

\section{SOC dynamical localization: nonlinear case}

In the presence of contact interactions the spectrum cannot be computed analytically but it can be computed numerically, with high accuracy,  by means of  self-consistent diagonalization (SCD) procedure \cite{salerno}. Quite surprisingly, we find that the  results of the previous section for  DL points survive  in the presence of nonlinearity. This is shown in
Fig.\ref{fig3}  where the spectrum of the Hamiltonian versus $\eta$ is depicted for the cases of all attractive interactions (top panel) and all repulsive interactions (bottom panel).
We see that, except for the presence of additional non degenerate bound state curves introduced by the interactions, both in the semi-infinite and interband gaps, the top and bottom bands curves are very similar to the ones depicted in Fig.\ref{fig2},  shrinking  into single points and  leading to flat bands in $k$-space, at exactly the same values of $\eta$ derived in Eq.(\ref{fbseq}) for the linear case (similar results are found for other values of the nonlinearity parameters).
The fact that the DL points are unaffected by the interaction is a consequence of the dimer localization and of the onsite nonlinearity.

From Fig.\ref{fig3} it is also clear that  while in the attractive case the degeneracy of the ground state at the DL points is fully removed by the nonlinearity,  in the repulsive case the ground state remains highly degenerated and almost unaffected by the interaction  (see bottom panel of Fig.\ref{fig3}). We remark that, in analogy with gap solitons of the continuous Gross-Pitaevskii equation \cite{KS2002}, the non degenerate localized states which appear in the band gaps
originate from  linear Bloch states at the center (resp. edges) of the Brillouin zone that become modulationally unstable when an attractive (resp. repulsive) interaction is switched on. The energies (chemical potentials) of these states, due to their  negative (resp. positive) interaction energy contributions, are pulled just below (resp. above)  the linear flat band value in the case of attractive (resp. repulsive)  interactions. This explains why the ground state  degeneracy at the DL points is fully removed for attractive interactions but not for repulsive interactions. The irrelevance of repulsive nonlinearities for flat band ground states also correlates with similar behaviors observed in the pure quantum regime of flat band interacting bosons in the  small density limit\cite{quantumflatbands}.

\begin{figure}[t]
\vspace{0.5cm}
\begin{center}
\includegraphics[width=6cm]{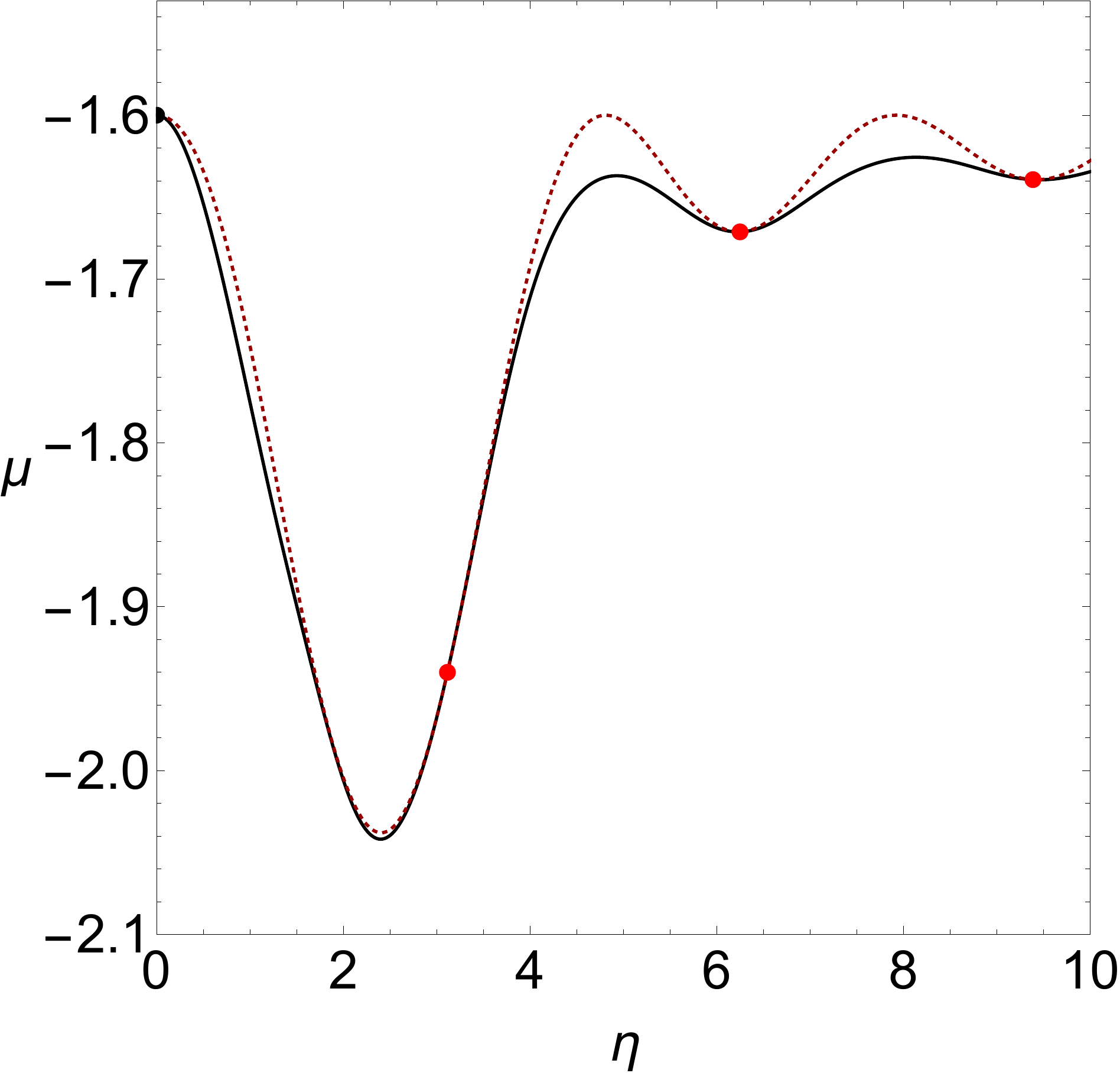}
\vspace{-0.cm}
\end{center}
\caption{Comparison between the exact nonlinear ground state curve in Fig. \ref{fig3} (black solid line) and the corresponding approximated one obtained from Eq. (\ref{nonlinearcase}) (red dotted line). Notice that the two curves coincide at the DL points (red dots) and are in very good agreement both around these points and in the  range  $0<\eta \approx 4$. All parameters are the same as in Fig. \ref{fig3}.
}
\label{fig4}
\end{figure}

In the following we restrict  to the case of all attractive interactions and concentrate on the ground state curve shown in the top panel of Fig. \ref{fig3}. In this respect we remark that the ground states at the DL points can be computed exactly by assuming the same type of dimerized localization found in the linear case. In this respect let us  look for states localized on two sites of the form $U_{n_0}=a e^{-i \epsilon t}$ , $V_{n_0+1}=i b e^{-i \epsilon t}$ with a, b real constants and with $U_{n}=0$, and   $V_{n}=0$ on all other sites different from $n_0$ and $n_0+1$, respectively.
Substituting into Eq. (\ref{avGP})  and looking for stationary solutions, we obtain the following cubic system:
\begin{eqnarray}
\frac 12 \sigma J_0^+ b - (\mu - \Omega_0)) a + \gamma_1 a^3=0,  \\
\frac 12 \sigma J_0^+ a - (\mu + \Omega_0)) b + \gamma_2 b^3=0,
\label{nonlinearcase}
\end{eqnarray}
which can be solved, together with the normalization condition $a^2+b^2=1$, exactly for $a,b,\mu$, with $\mu=\epsilon+E_{int}$ denoting the chemical potential.
Here we assumed $\eta$ to be a solution of the equation $J_0^-=0$, but results for the  case  $J_0^+=0$ follows in similar manner. From this  we obtain exact  nonlinear localized states at the DL points  and, although the analytical expressions for $a, b$ are too involved to be reported, it is possible to obtain them  numerically with high accuracy. Despite these results are strictly valid at the DL points where bands are flat, the above equations can be solved in general by considering $\eta$ a varying parameter, to see how results deviate (away from DL points)  from the ones obtained by SCD  in Fig.\ref{fig3}. This is shown in Fig.\ref{fig4} for the ground state energy curve. Quite remarkably, we see that the comparison with the SCD curve in Fig.\ref{fig3} is exact at the DL points and  very good for a wide interval, at least up to  $\eta \approx 4$.

\section{Discussion and conclusions}
We  briefly discuss possible  parameters design and experimental setting to observe
the above results presented in this paper. In this respect we refer to the SOC for
the case of $^{87}$Rb atoms in the field of three laser beams
implemented in a tripod scheme. The ground states from the
$5S_{1/2}$ manifold are coupled via differently polarized light, by
choosing $|1\rangle = |F = 2,m_F = -1 \rangle$, $|2\rangle = |F = 2,m_F = +1\rangle$,
and $|3\rangle = |F = 1,m_F = 0\rangle$ \cite{edmonds}.

The optical lattice can  be generated  with two additional counterpropagating linearly polarized laser beams
of wavelength $\lambda=2 \pi/k_L=842$nm with a strength of the order of $10 E_R$ with $E_R$ the recoil energy $E_R=\hbar^2 k^2/(2 m)$. These values guarantee the  applicability of the tight-binding model we used. The passage of the optical lattice laser beams through an acousto-optic modulator permits one to  introduce a frequency difference between them which can be used for the shaking of the lattice as discussed in Ref. \cite{arimondo}.
The strong modulation limit can be reached by considering
a Zeeman field of normalized amplitude $>20$ and
frequency of the modulation fixed by $\omega= 2 \Omega_1/\eta$.
Under these circumstances, it should be possible to observe the described results and in particular the dimer localization properties  of the ground state at
the DL points.

In conclusion, we have discussed flat bands, tunneling suppression and  DL in binary BEC mixtures with spin-orbit coupling subjected to a shaking optical lattice and periodic time modulations of the Zeeman field. For this we have used a tight binding model of the BEC mixture valid for deep optical lattices
and considered the effects of the modulations by means of the averaging method.
In particular, we  showed that the suppression of the interwell tunneling is not enough to observe the DL phenomenon, and a suitable tuning of  the SOC parameter  with the  optical lattice shaking is required.
The SOC tuning, achieved via the Zeeman field, was shown to lead to a series of parameter values  at which flat bands and DL can occur.

Exact analytical expressions of  the BEC wave functions at the DL points have been derived, both in absence (linear case) and and in presence  (nonlinear case) of interactions. In the latter case we have shown that the dimer localization occurs also for the ground state of the system, in exact form, at DL points, and in a very good approximation, around these points.
Parameters design for possible experimental implementations of the above phenomena were also briefly  discussed.
\vspace{0.2cm}

\section*{Acknowledgments}
M.S. acknowledges partial support from the Ministero dell'Istruzione,
dell'Universit\'a e della Ricerca (MIUR) through the PRIN (Programmi di
Ricerca Scientifica di Rilevante Interesse Nazionale) grant on
"Statistical Mechanics and Complexity": PRIN-2015-K7KK8L.

\end{document}